\documentclass[12pt]{iopart}
 
 \usepackage  {subfigure}
 \usepackage{graphicx}

\begin{document}
%\preprint{}

\title[Quantum phase transition in  FK model]{Quantum phase transition in one-dimensional commensurate
  Frenkel-Kontorova model}

\author{Yongjun Ma$^1$, Jiaxiang Wang$^1$, Xinye Xu$^1$, Qi Wei$^1$ and Sabre Kais$^2$$^3$}
\address{$^1$State Key Laboratory of Precision Spectroscopy and
Department of Physics, East China Normal University, Shanghai
200062, China}
\address{$^2$Departments of Chemistry and Physics, Purdue University,
West Lafayette, Indiana 47907, USA}
\address{$^3$Qatar Environment and Energy Research Institute, Qatar Foundation, Doha, Qatar}

\ead{ jxwang@phy.ecnu.edu.cn}
\begin{abstract}
  In this paper, we have studied the one-dimensional
commensurate quantum Frenkel-Kontorova model by a density-matrix
renormalization group (DMRG) algorithm. The focus has been on its
properties of the entanglement, the coordinate correlation, the
ground state energy and the energy gap between the ground state and
the first excited one. It is demonstrated that a quantum phase
transition (QPT) between the pinned and the sliding phases takes
place as the quantum fluctuation, measured by an effective Planck
constant $\tilde\hbar$, increases to a threshold value $\tilde \hbar_c$.

\end{abstract}
\pacs{05.10.Cc,64.70.Tg}
%\noindent{\it Keywords}:Frenkel-Kontorova model, quantum phase transition, entanglement

\maketitle
\section{Introduction}
\label{intro}
The Frenkel-Kontorova model (FK model) \cite{FK1,FK2,FK3} describes
a chain of interacted particles in the presence of an external
periodic potential which has attracted much attention ever since it
first appeared in last century \cite{F-K model1}. In the continuum
approximation, the model will be reduced  to the well-known
integrable sine-Gordon equation. But as a discrete model, it is nonintegrable and has
been a generic tool to study many nonlinear effect such as chaos,
kinks, breathers and so on \cite{F-K model1,F-K
model2,F-K model3,F-K model4}. In condense matter physics, it also has a wide range of application,
such as describing an adsorbate layer on the surface of a crystal \cite{aubry1978},
charge-density-wave transport \cite{floria}, dry friction \cite{Braun1997} and a chain
of coupled Josephson junctions \cite{Watanabe}.
\par
Many of the interesting properties of the classical FK model come from its two competing length scales:  the average distance
between the particles and the  length of the spacial period of the
external potential. For example, if the ratio of the
two scales is a rational number, the system is commensurate,
otherwise, it is incommensurate. For the former case, a classical
FK model is always in a pinned state. For the latter case, there is
a threshold $K_c$.  If $K>K_c$, the particles are still pinned.
But if $K<K_c$, the particles are depinned, and  can slide along the
external potential.
\par
No matter whether the classical is commensurate or incommensurate,
intuitively, all the pinned state can transit to a sliding one so
long as enough quantum fluctuations are introduced. Up to now, there
have been a lot of discussions upon this quantum phase transition
(QPT) for the incommensurate case \cite{F-K5,F-K6,F-K7,F-K8}. But as
far as QPT is concerned, the commensurate case is also a very
interesting topic to discuss since it is a physically more clean
case and we will be save from all those classical complexities
specific to incommensurate FK model any more.

Unlike the classical phase transition during which temperature plays
a major role, QPT occurs as a result of the quantum fluctuations. In
condensed matter physics, QPT is a very hot subject due to its close
connection to some fundamental problems such as quantum criticality
and high-temperature superconductivity \cite{sachdev2011}. Usually,
the QPT is driven by changing some nonthermal external parameters.
But for the QPT we are going to study in this paper, it is induced
directly by the quantum fluctuation just like we change the
temperature in classical phase transition. It is hoped that this
different angle can provide some new perspective to QPT since the
physical detail of the QPT process is still far from being well
understood.
\par
 In the article, we will try to use the entanglement, the coordinate
correlation, the ground state energy and the energy gap
between the ground state and the first excited state to find the
transition point.
\par
To solve the quantum FK model, the density matrix renormalization
group (DMRG) method will be used, which is developed more than two
decades ago by White\cite{dmrg01}. It is well known for the
high-accuracyin simulating the low-dimensional strongly correlated
quantum systems \cite{dmrg2}. The scheme of standard DMRG algorithm
can be found in many reference \cite{dmrg0,dmrg01,dmrg1,dmrg2}
and the concrete realization in FK model has been given in Refs.
\cite{F-K4}. So the the detail will not be presented here. We will
just briefly review the FK model to introduce the necessary
notations in the next section.
\section{Quantum FK Model}
\begin{figure}[htbp]
\centering
\includegraphics[angle=-90,clip=true,width=14cm]{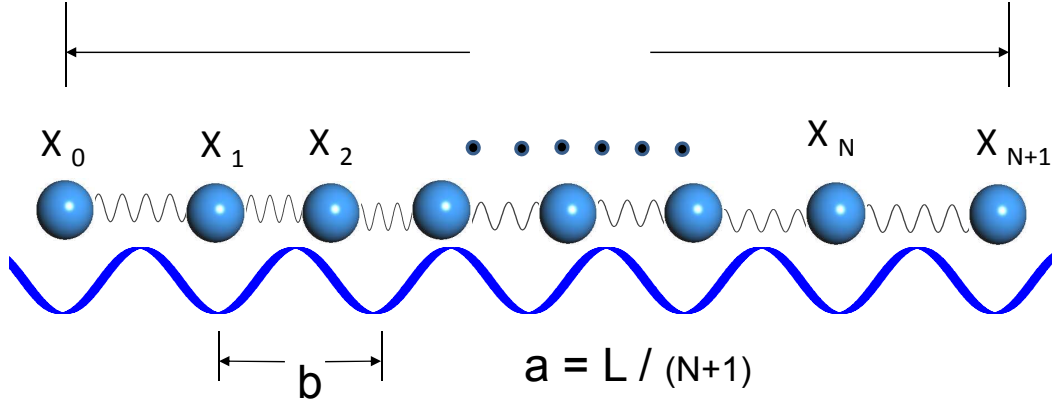}
\caption{ Schematic diagram of the Frenkel-Kontorova model. A chain
of $N+2$ particles connect by springs  is subjected to the action of
external sinusoidal potential with spacial period $b$. The average
distance between neighboring particles is $a=L/(N+1)$.}\label{model}
\end{figure}
Fig. \ref{model} presents the schematic diagram of the
one-dimensional Frenkel-Kontorova model \cite{F-K model1}: a chain
of $N+2$ particles connected by springs  and subjected to the action
of an external sinusoidal potential. The average distance between
neighboring particles and the external potential period are
$a=L/(N+1)$ and $b$, respectively, where $L$ is the length of the
chain.  For the commensurate case we are interesting in this paper,
$a/b$ is set to be $1$. By using the fixed boundary conditions with
$x_{0}=0, x_{N+1}=(N+1)a$, the Hamiltonian can be expressed as,
\begin{equation}
\hat{H}=\sum_{i=1}^{N}[- \frac{\hbar^2}{2m}\frac{\partial^2
}{\partial x_{i}^{2}}+ \frac{k}{2}( \hat x_{i+1} - \hat x_{i} -a)^2
-V\cos(\frac{2\pi}{b} \hat x_{i} )],     \label{E_interna0}
\end{equation}
where $\hbar$ is the Planck, $k$ is the elastic
constant and $V$ is the potential amplitude.
  \par
With $q_{0}=2\pi/b$ , we introduce the dimensionless
parameters,
\begin{eqnarray}
     && \hat X_{i}  =  q_{0}{\hat x_{i}}, \quad \hat U_{i}=\hat X_{i}-i\cdot2\pi,\quad
  \hat{H'}=\frac{q_{0}^{2}}{k}\hat{H} ,\nonumber\\
  &&K=V\frac{q_{0}^{2}}{k},\quad  \tilde{\hbar}=\frac{q_{0}^{2}}{\sqrt{mk}}\hbar.  \label{daihuan}
\end{eqnarray}

 Then the Hamiltonian can be rewritten in a dimensionless form,
\begin{equation}
\hat{H'}=\sum_{i=1}^{N}[- \frac{ \tilde{\hbar}^2}{2}\frac{\partial^2
}{\partial U_{i}^{2}}+ \frac{1}{2}({\hat U_{i+1}}-{\hat U_{i}})^2
-K\cos(\hat{U_{i}}+i\cdot2\pi)] \label{reaction2}.
\end{equation}
By using the following transformations,
\begin{equation}
  \hat{U_{i}} = \frac{1}{\sqrt{2}} \frac{\sqrt{ \tilde{\hbar} }}
{\sqrt[4]{2}}   ({\hat a_{i}}^{\dag}+{\hat a_{i}}),
\nonumber
\end{equation}
\begin{equation}
\hat{P_{i}}= \frac{\partial }{\partial{U_{i}}} = -\frac{1}{\sqrt{2}}
\frac {\sqrt[4]{2}}{\sqrt{ \tilde{\hbar} }} ({\hat a_{i}}^{\dag}-{\hat
a_{i}}),
\end{equation}
where $\hat a_{i} ^{\dag}$ and $\hat a_{i} $ are the annihilation
 and creation operators satisfying $[\hat a_{i} ,\hat a_{j}^{\dag} ]=
 \delta_{ij}$,  $[\hat a_{i} ,\hat a_{j} ]= 0$,
 $[\hat a_{i} ^{\dag},\hat a_{j}^{\dag} ]= 0$,
we can rewrite the Eq.\ (\ref{reaction2}) into a
 second-quantized form,
\begin{eqnarray}
\hat{H'}&=& \sqrt{2} \tilde{\hbar}  \sum_{i=1}^{N} \{({\hat
a_{i}}^{\dag}{\hat a_{i}}+\frac{1}{2} )   \nonumber\\
 &&-\sqrt{2} \tilde{\hbar} \sum_{i=1}^{N}(\hat a_{i} ^{\dag} +\hat a_{i})(\hat a_{i+1}^{\dag} +\hat a_{i+1})\nonumber\\
&& -\frac{K}{\sqrt{2} \tilde{\hbar} }\cos[\frac{1}{\sqrt{2}} \frac{\sqrt{ \tilde{\hbar}} }
{\sqrt[4]{2}} (\hat a_{i} ^{\dag}+\hat a_{i} )+i\cdot2\pi]\}.
   \label{action4}
\end{eqnarray}
From the above equation, it is easy to see that there are two independent parameters:
$\tilde{\hbar} $ and $K$.   The first is the effective Plank
constant, denoting the quantum fluctuation. The second
measures the strength of external potential. These two free effective
parameters can be varied by changing the elastic constant $k$,
the particle mass $m$, the spacial period $b$ or the magnitude of the external
potential $V$ according to Eq. (\ref{daihuan}).

\section{RESULTS AND DISCUSSIONS}
In the incommensurate case, some dumb particles (or glue
particles) have been found in FK model, which are
located very close to the bottoms of the potential, dividing  the
whole chain into several subblocks (bricks) \cite{F-K11}. The particles inside the
subblocks can interact each easily and remains unaffected by the
outside particles. As the quantum fluctuation increases, the dumb
particles will change into normal ones and the bricks melt, thus a
new sliding phase appears and the QPT occurs. In the commensurate
case, the same scenario can be realized since every particle will look like a
dumb particle. Then as the quantum fluctuation is introduced, QPT will happens.

\subsection {Entanglement}
\begin{figure}[htbp]\centering\subfigure[ ]{\label{fig:entangle1:a}
\includegraphics[angle=-90,width=10cm]{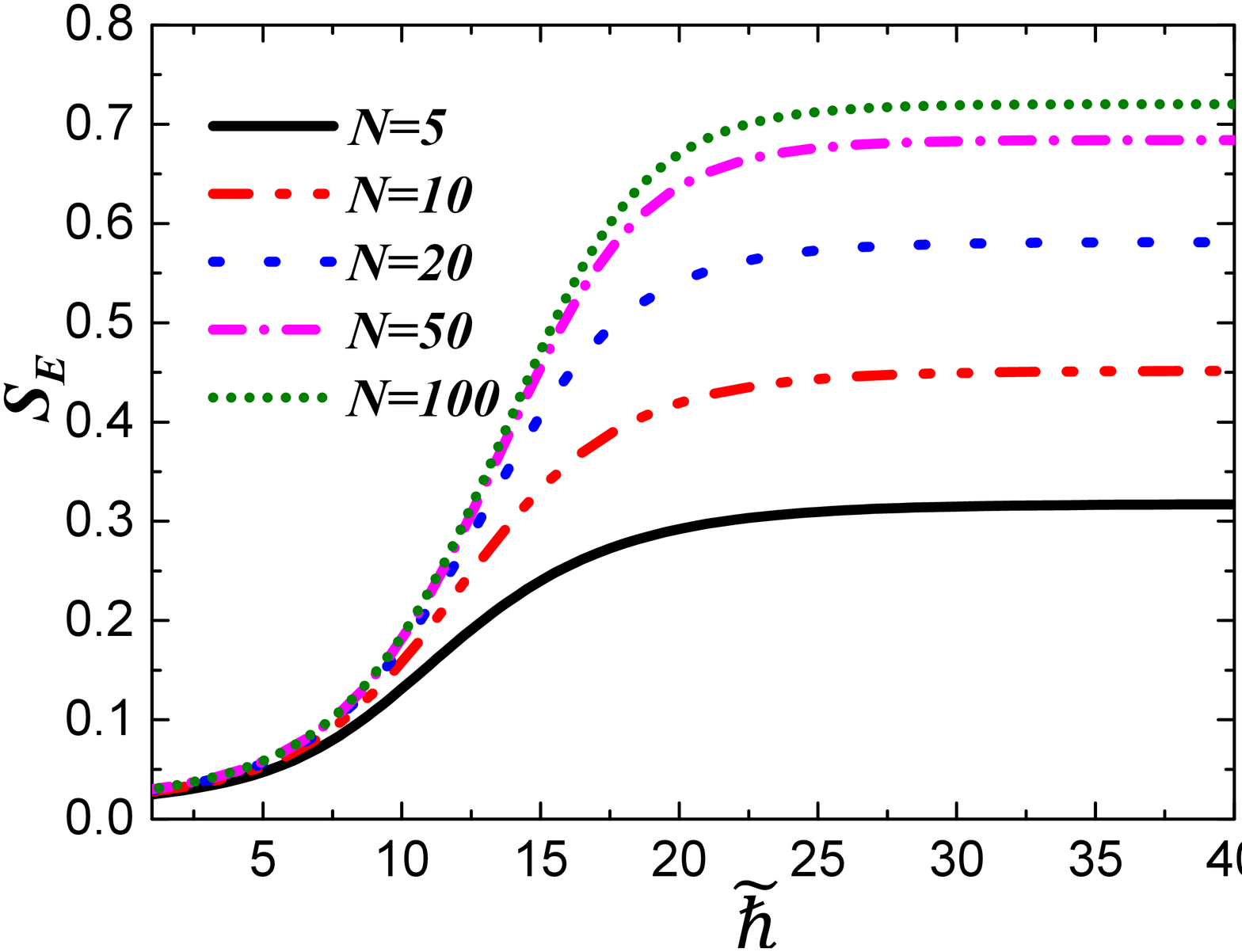}}
\subfigure[ ]{\label{fig:entangle1:b}
\includegraphics[angle=-90,width=10cm]{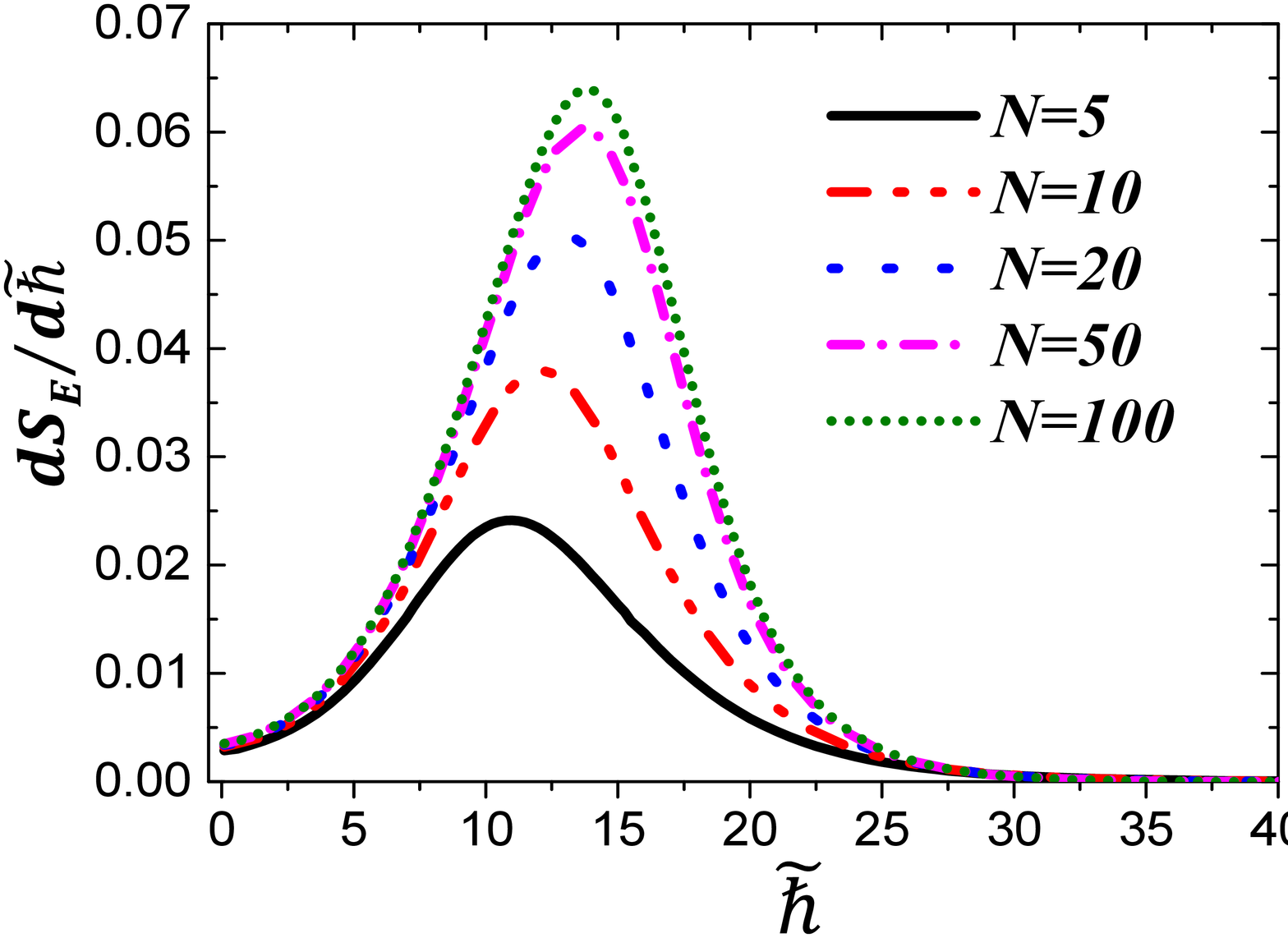}}
 \caption{(a)The variation of entanglement $ S_{E}$ against the
effective Planck constant $\tilde{\hbar}$ for different system size with $K
= 5$. (b) The same as in (a) but for the  first-order derivative of
$S_E$.}\label{entangle1}

\end{figure}

As a complex quantum phenomenon, the entanglement is famous for its
possible utilization as a resource in quantum communications and
computations \cite{Kais}. Recent work has shown that entanglement
can also be used as an order parameter to study QPT in condensed
matter physics. For example, it has been demonstrated that for many
one-dimensional magnetic systems, the entanglement shows scaling
behaviour in the vicinity of the transition point \cite{entangle0}.
For the moment, a lot of  work have been done upon the relation
between the entanglement and QPT
\cite{entangle1,entangle2,entangle3,entangle4,entangle5,entangle6}.
Hence we have a good reason to check the entanglement in our work.

The von Neumann entropy will be used to measure the entanglement
\cite{entangle7}, which is defined as,
\begin{equation}
S=-tr(\rho\log \rho ),
\end{equation}
where $\rho$ is the density matrix of the state we are interested in
and $tr$ denotes the trace In our work, the following average
entanglement between a single particle and the rest of the systems
is calculated,
\begin{equation}\label{7}
  S_{E}=\frac{1}{n} \sum\limits_{i=1}^n S_i,
\end{equation}
in which $S_i=-tr_i(\rho_i \log\rho_i)$ is the von Neumann entropy  corresponding to $i$th
site with $\rho_i=Tr_i\left |\Psi\right\rangle\left\langle \Psi\right|$\cite{entangle1,entangle2,entangle3}. Here $Tr_i$ stands for
the tracing over all sites except the $i$th site.

\par
Fig. \ref{entangle1}(a) gives the variation of $S_{E}$ against
$\tilde{\hbar}$ for different system size with fixed external potential
$K=5$. It is easy to see that $S_E$ increases with $\tilde{\hbar}$ until it
reaches the maximum. Moreover, for bigger system size, $S_E$
increases faster with $\tilde{\hbar}$, but reaches the maximum value more
slowly. This implicates that although the smaller system has lower
maximum entanglement, as a whole it can be entangled much easier
when the quantum fluctuation is introduced. In other words, we need
higher quantum fluctuations to entangle the bigger system. This is
consistent with our physical intuitive. As the system size become
bigger and bigger, it is important to note that the curves tend to
converge to one curve. To see the significance of this effect, we
plot the first-order derivative of $S_E$ in Fig. \ref{entangle1}(b).
One marvelous observation is that there is a peak for each curve.
And the position of the peak approaches a definite value of
$\tilde{\hbar}_c\approx13$ as the system size is increased. From the recent
research upon the entanglement and QPT \cite{entangle6}, we can
regard $\tilde{\hbar}_c$ here as a phase transition point from pinned state
to a sliding one. Compared with the transition point at
$\tilde{\hbar}_c\approx 2$ for the incommensurate quantum FK model with the
same external potential \cite{F-K4,F-K5}, $\tilde{\hbar}_c\approx13$ is a
much higher value. This difference can be attributed to their
classical behaviours. For example, the incommensurate FK system can
only get pinned when $K$ is bigger than $1$, which is called
Transition by Breaking Analyticity (TBA) in classical physics
\cite{aubry1978}. But the commensurate FK system will get pinned for
any slight of external potential.
\par

\begin{figure}[htbp]\centering\subfigure[ ]{\label{fig:phase:a}
\includegraphics[angle=-90,width=10cm]{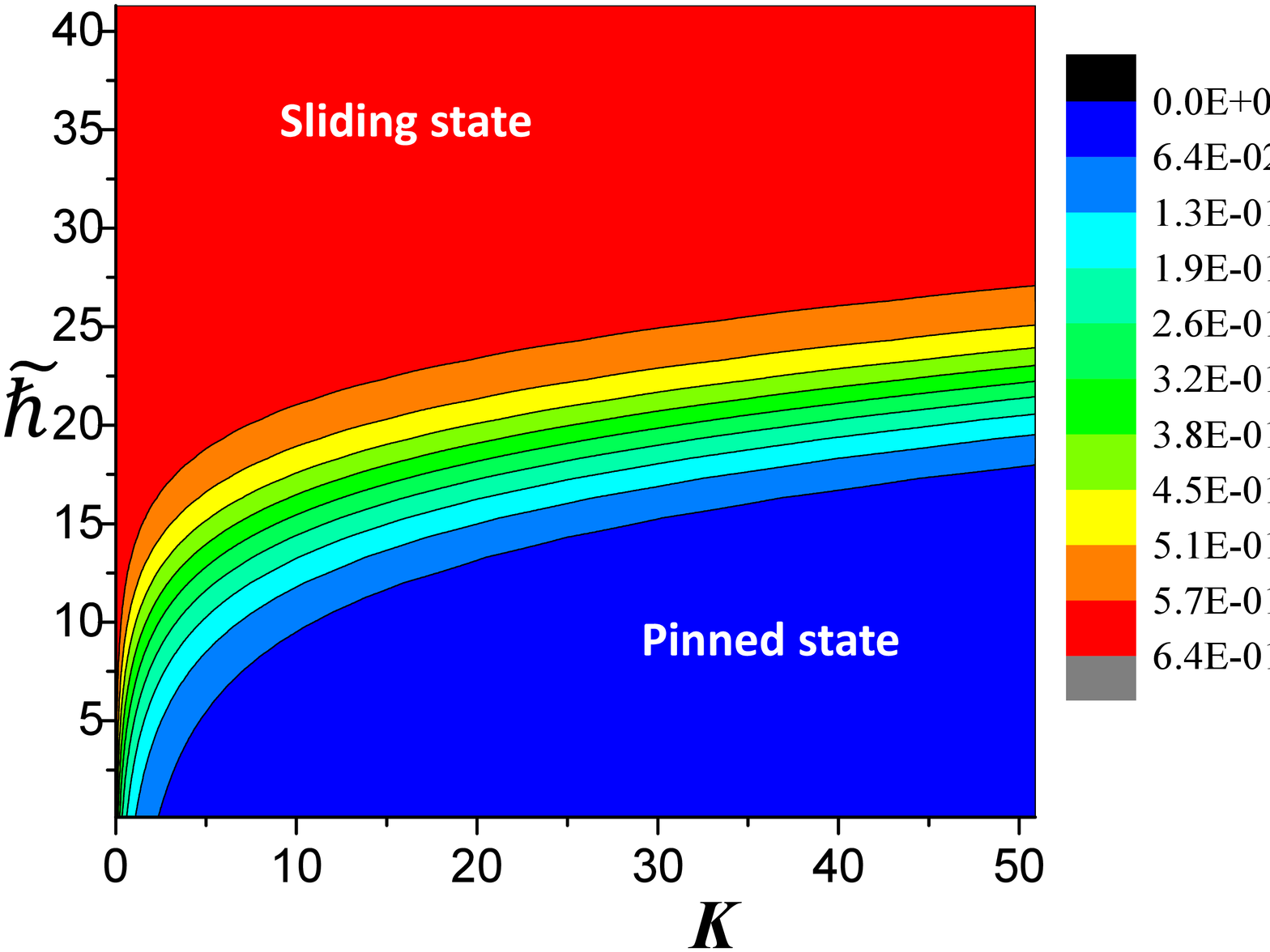}}
\subfigure[ ]{\label{fig:phase:b}
\includegraphics[angle=-90,width=10cm]{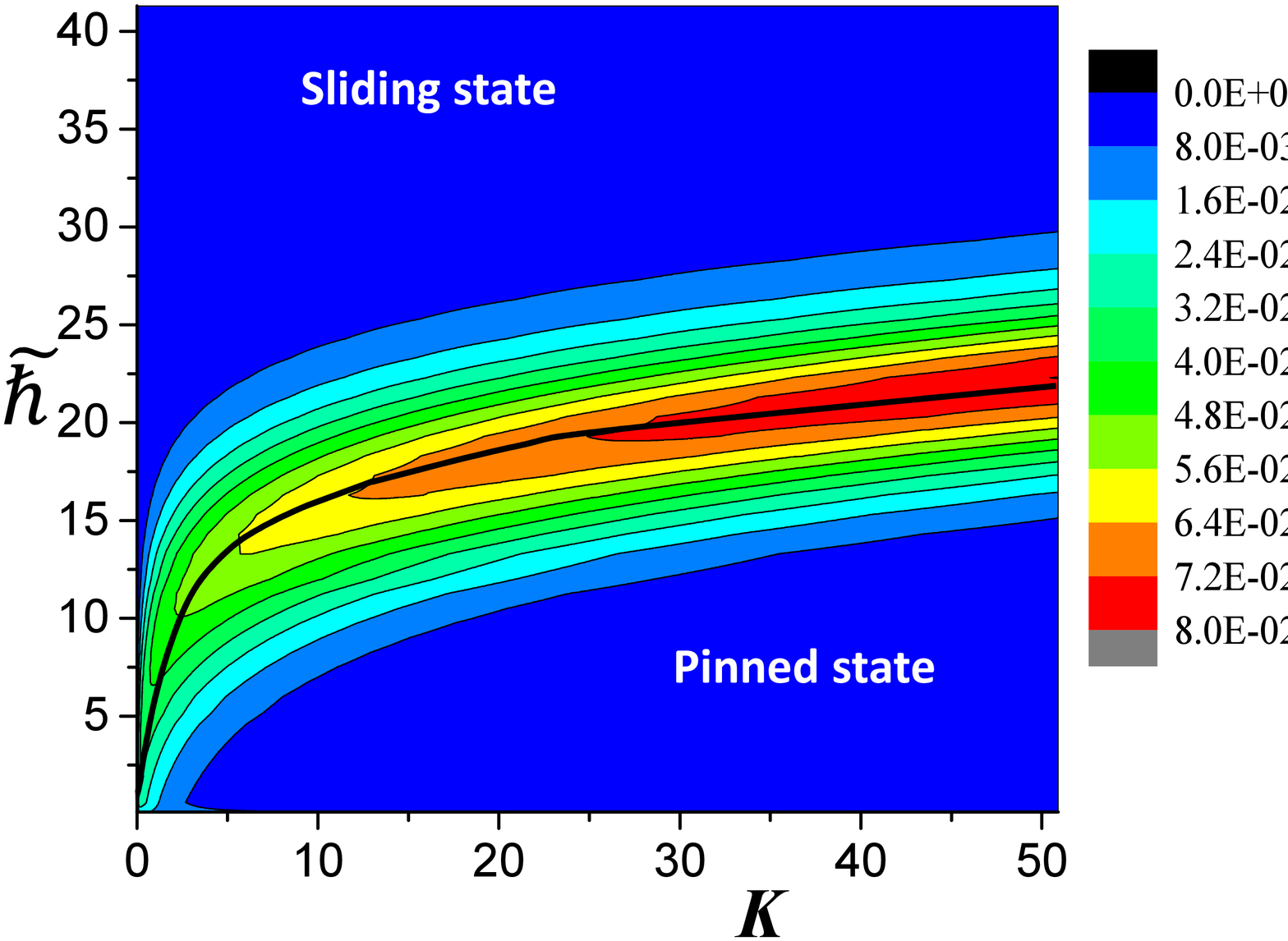}}
 \caption{ Contour map demonstrating the dependence of $S_E$ in (a)
and $\frac{d S_E}{d \tilde{\hbar}}$ in (b) upon $\tilde{\hbar}$ and $K$,
respectively. $N=50$. The skeleton of maximum points give the phase transition line(the black line in(b)).}\label{phase}
\end{figure}

For different external potential, the transition point of $\tilde{\hbar}_c$
will change. In Fig. \ref{phase}(a), we presents the contour map of
$S_E$ in the $K-\tilde{\hbar}$ parameter space. Although $S_E$ is a
complicated function of both $K$ and $\tilde{\hbar}$, we can still see two
obvious phases. The pinned phase is in the region where $K$ is large
or $\tilde{\hbar}$ is small. The sliding phase is in the region where $K$ is
small or $\tilde{\hbar}$ is large. There exists a transition zone between
the two phases, which illustrates how the transition happens. In
order to find out the transition points, we also give the contour
map of $\frac{d S_E}{d \tilde{\hbar}}$ on the $ \tilde{\hbar} - K $ plane
in Fig. \ref{phase}(b). The transition curve can be obtained by
tracing the points of the maximum $\frac{dS_E}{d \tilde{\hbar}}$ for each
definite $K$ in the direction of $\tilde{\hbar}$ axis or just roughly by
following the tips of the contour lines. For example, when $K=5$, $\tilde \hbar_c\approx 13$ can be acquired, which is consistent with the
result given by Fig. \ref{entangle1}(b).

\subsection {Coordinate correlation}

\begin{figure}[htbp]
\centering\subfigure[ ]{\label{fig:cl:a}
\includegraphics[angle=-90,width=10cm]{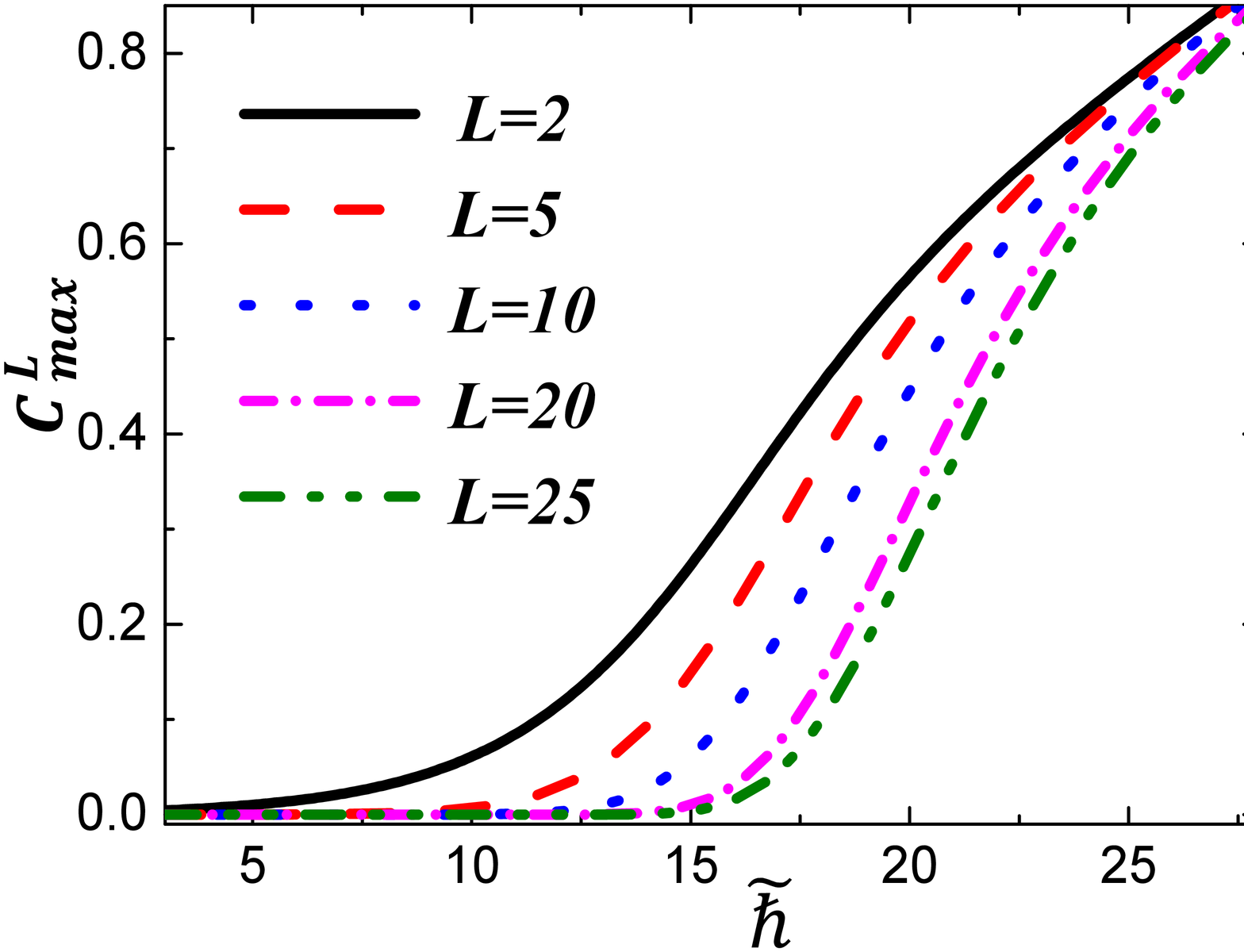}}
\subfigure[ ]{\label{fig:cl:b}
\includegraphics[angle=-90,width=10cm]{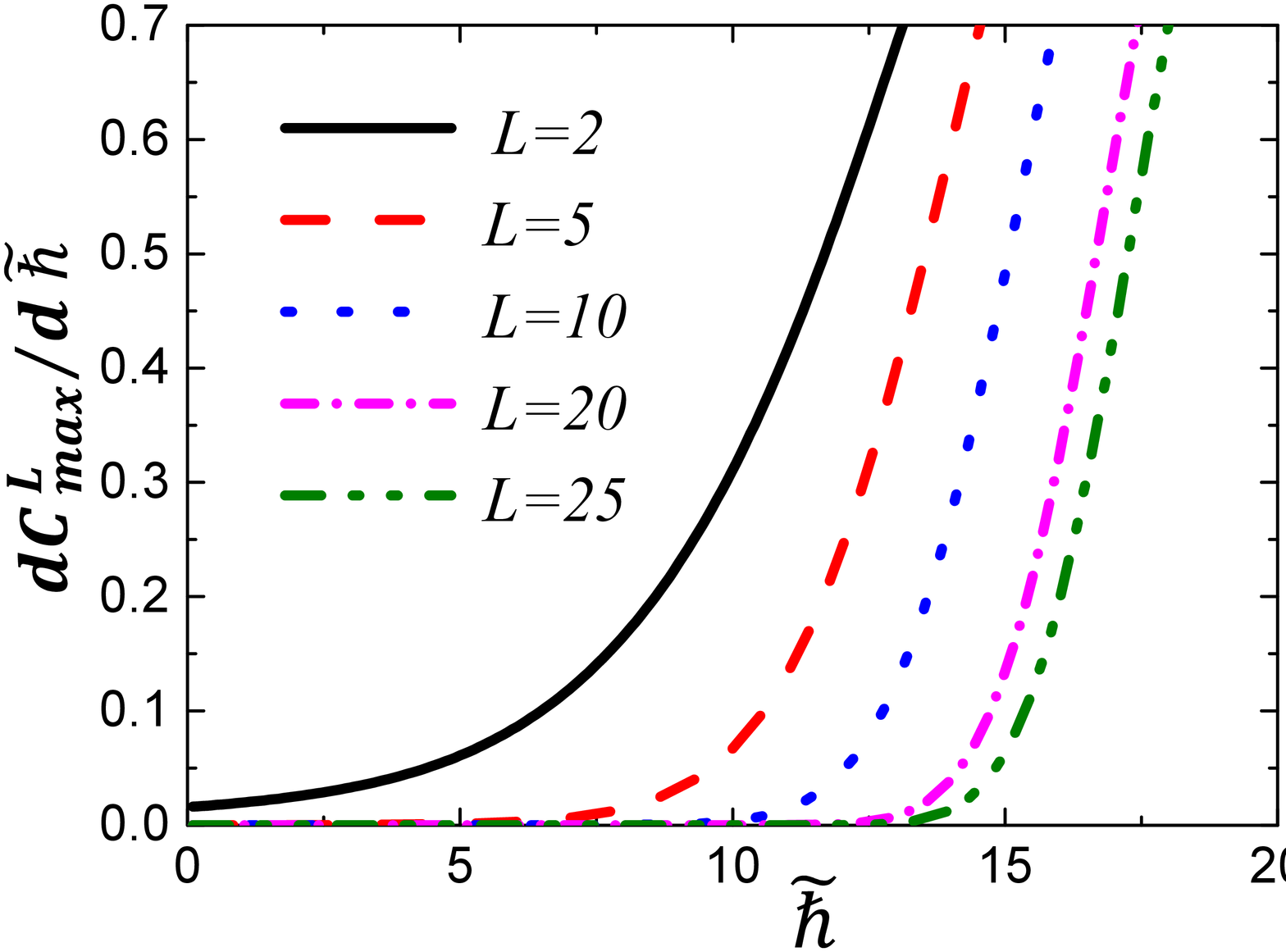}}
 \caption{(a) Variation of the correlation of different lengthes
denoted by $L$ against the quantum fluctuation. The system size is
set to $N=50$ and external potential $K=5$. (b) The same as in (a)
but for the first-order derivative of $S_E$ with respect to $\tilde{\hbar}$.}\label{cl}
\end{figure}

Besides the entanglement, there is also other kind of quantum
correlations.  In this section, we will discuss the coordinate
correlation, which is written as
\begin{eqnarray}\label{8}
\tilde{\hbar}  G_{ij}=<\Psi_0\mid(X_i-\bar{X}_i)(X_j-\bar{X}_j)\mid\Psi_0>,
\end{eqnarray}
from which we can define the  correlation of length $L$,
\begin{equation}
C^L=\tilde{\hbar} G_{ij}\delta( \left | i-j\right |- L).
\end{equation}
For $i,j$ satisfying $i-j=L$, the maximum value of $C^L$ will be
used to embody the coordinate correlation of length $L$, denoted as
$C^L_{max}$. Fig. \ref{cl}(a) gives the relationship between $C^L_{max}$
and $\tilde{\hbar}$ for different correlation length $L$. For comparison,
$C^L_{max}$ has been normalized [0,1] for each curve and the system size
is chosen to be $50$. It is interesting to see that as the quantum
fluctuation begins to set in, the correlation between the particles
with distance $L=2$ emerges first and increases with $\tilde{\hbar}$. Then
correlation with longer length $L$ begins to appear when
$ \tilde{\hbar}$ is increased further. Finally, around $\tilde{\hbar}_c\approx13$, the
correlation of length $L=25$, which is about half of system size $N=50$,
starts to dominate the system. This is also the transition point
corresponding to the entanglement that changes most rapidly with the
quantum fluctuation. Like entanglement, we have also studied the changing speed of $C^L_{max}$ with
$\tilde{\hbar}$, i.e. $\frac{d C^L_{max}} {d \tilde{\hbar}}$ , which is presented in Fig. \ref{cl}(b), where each curve is normalized to (0,1) just as in Fig. \ref{cl}(a).
The same point at $\tilde{\hbar}_c\approx 13$ can be found to correspond to the maximum increasing speed of the correlation with length $L=25$.
\par

\begin{figure}[htbp]
\centering
\includegraphics[angle=-90,width=10cm]{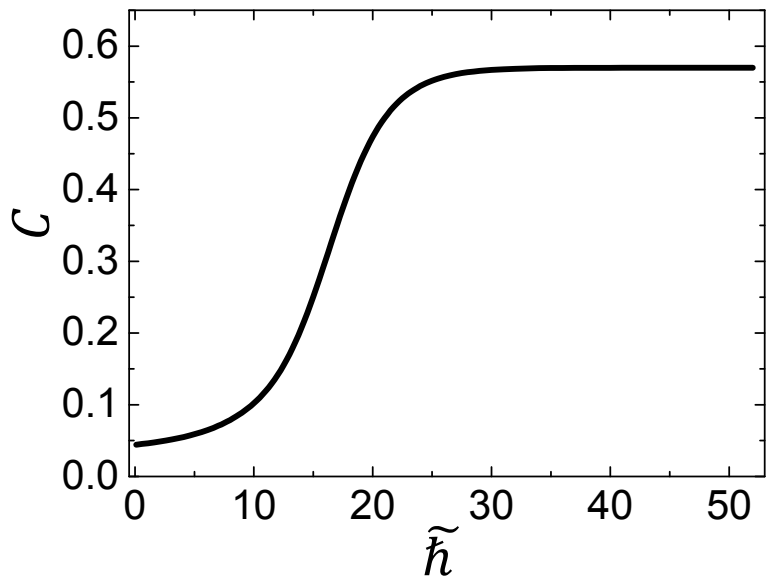} \caption{
  The entanglement $C$ against the effective Planck constant
$\tilde{\hbar}$ for a FK-chain  with $N=50$ and $K = 5$.   }
\label{co}
\end{figure}

In order to find out the relationship between the correlations and QPT,
we introduce the following parameter,
\begin{equation}\label{10}
C = \frac{\sum_{L=1}^N C^L_{FKmodel}}{~\sum_{L=1}^N C^L_{Harmomic}},
\end{equation}
where $C^L_{FKmodel}$ and $C^L_{Harmomic}$ denote the $C^L$ of the
FK model and the oscillator chain respectively. The results are given in
Fig. \ref{co}, from which we can observe an obvious change of the correlations around $\tilde{\hbar}_c\sim 13$. It is not a sharp change, which may result from the
finite-size effect.

\subsection{Ground state energy and energy gap}

For the ground state of a classical commensurate FK chain, the
particles are localized by the external potential. As the quantum
effect comes in, they will gain kinetic energy due to the quantum
fluctuations. As the gained kinetic energy is high enough to help
the particles overcoming the trapping potential, the whole system
can be expected to get depinned and become a sliding state.
\par
\begin{figure}[htbp]
\centering
\includegraphics[angle=-90,width=15cm,trim=-80 10 0 0]{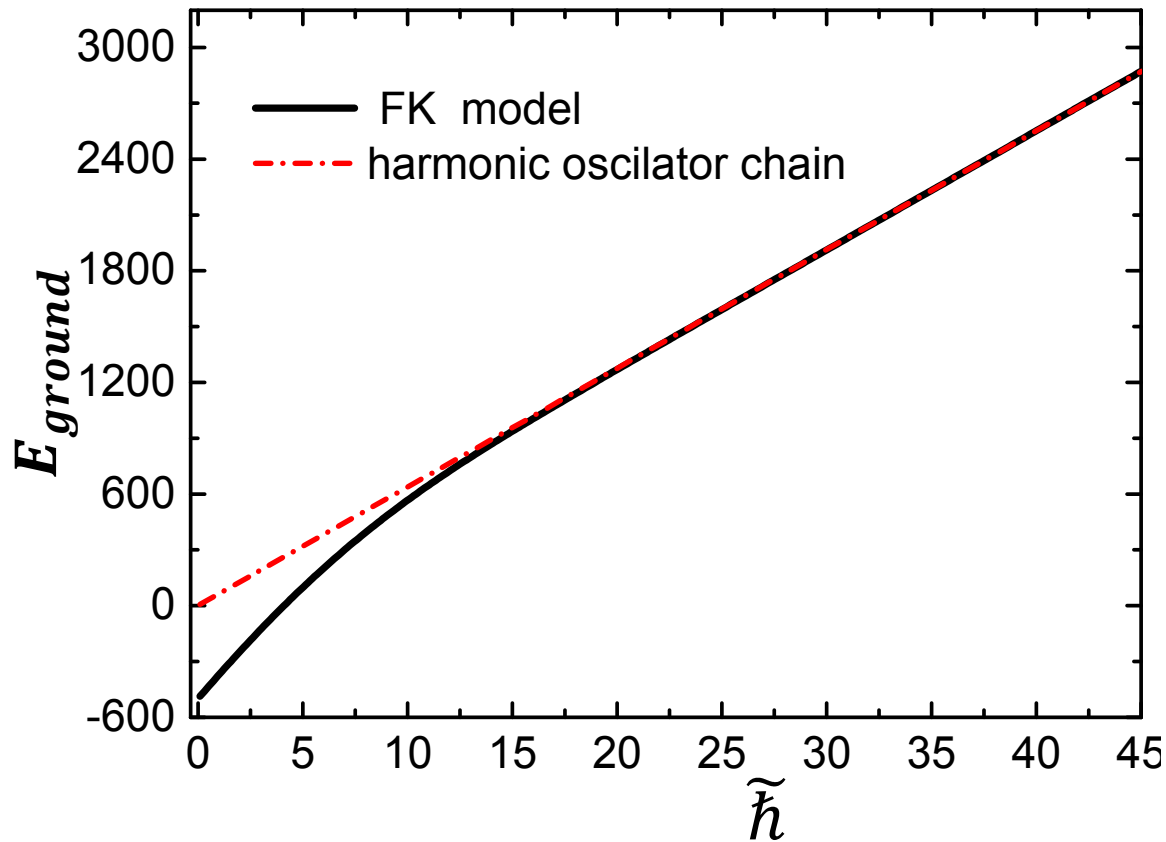}
\caption{ (a) The variation of the ground state energy against the
quantum fluctuation denoted by $\tilde{\hbar}$. For contrast, the results
for a chain of harmonic oscillators are also presented. (b)
gives the dependence of the energy gap $\Delta E_{12}$ and with respect to $\tilde{\hbar}$. The
other parameters in the calculations are $N=100$ and $K=5$. }
\label{energy}
\end{figure}
Fig. \ref{energy} gives the dependence of the ground state
energy $E_{ground}$ upon $\tilde{\hbar}$. For contrast, the corresponding
curve for a harmonic oscillator chain is also given. The chain size
is set to be $N=50$ for both systems and the external potential for
the FK model is $K=5$. It is apparent $E_{ground}$ increases with
$\tilde{\hbar}$. Moreover the difference of the ground state energy between
the FK model and the harmonic oscillator chain gets smaller and
smaller until the two curves almost overlap with each other at about
$\tilde{\hbar}_c\approx 13$, which implicates a phase transition  from the
pinned state to the sliding one. This phenomenon also tells us that
the behaviour of the sliding state is more like a harmonic
oscillator chain with the accretion of quantum fluctuations.

\begin{figure}[htbp]
\centering
\includegraphics[angle=-90,width=12cm,trim= 20 10 0 0]{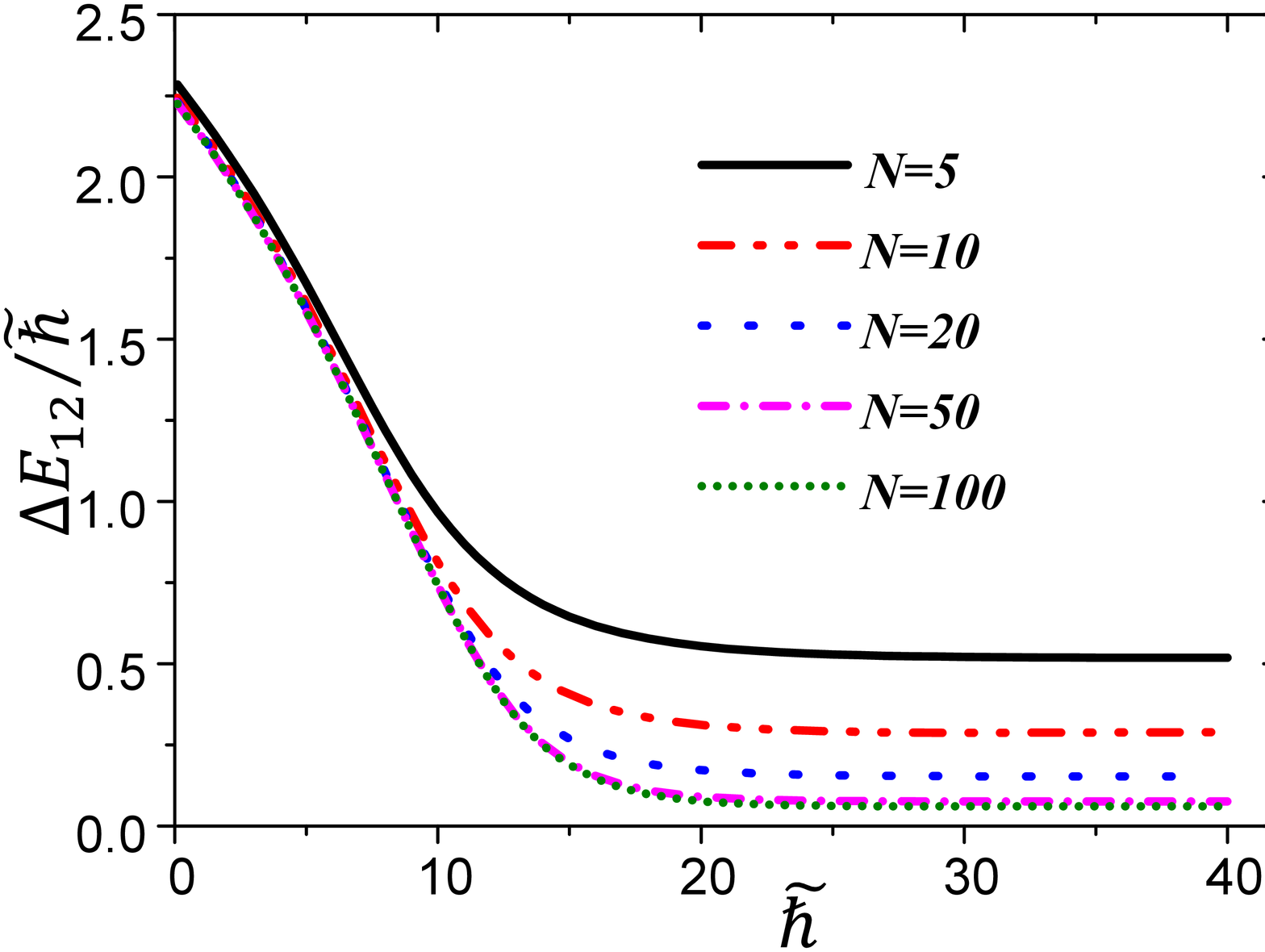}
\caption{   The variation of  the energy gap $\frac{\Delta E_{12}}{\tilde{\hbar}}$  against the
quantum fluctuation denoted by $\tilde{\hbar}$ for different system size. The
  parameters is $K=5$. }
\label{gap1}
\end{figure}

\begin{figure}[htbp]
\centering
\includegraphics[angle=-90,width=12cm,trim= 20 10 0 0]{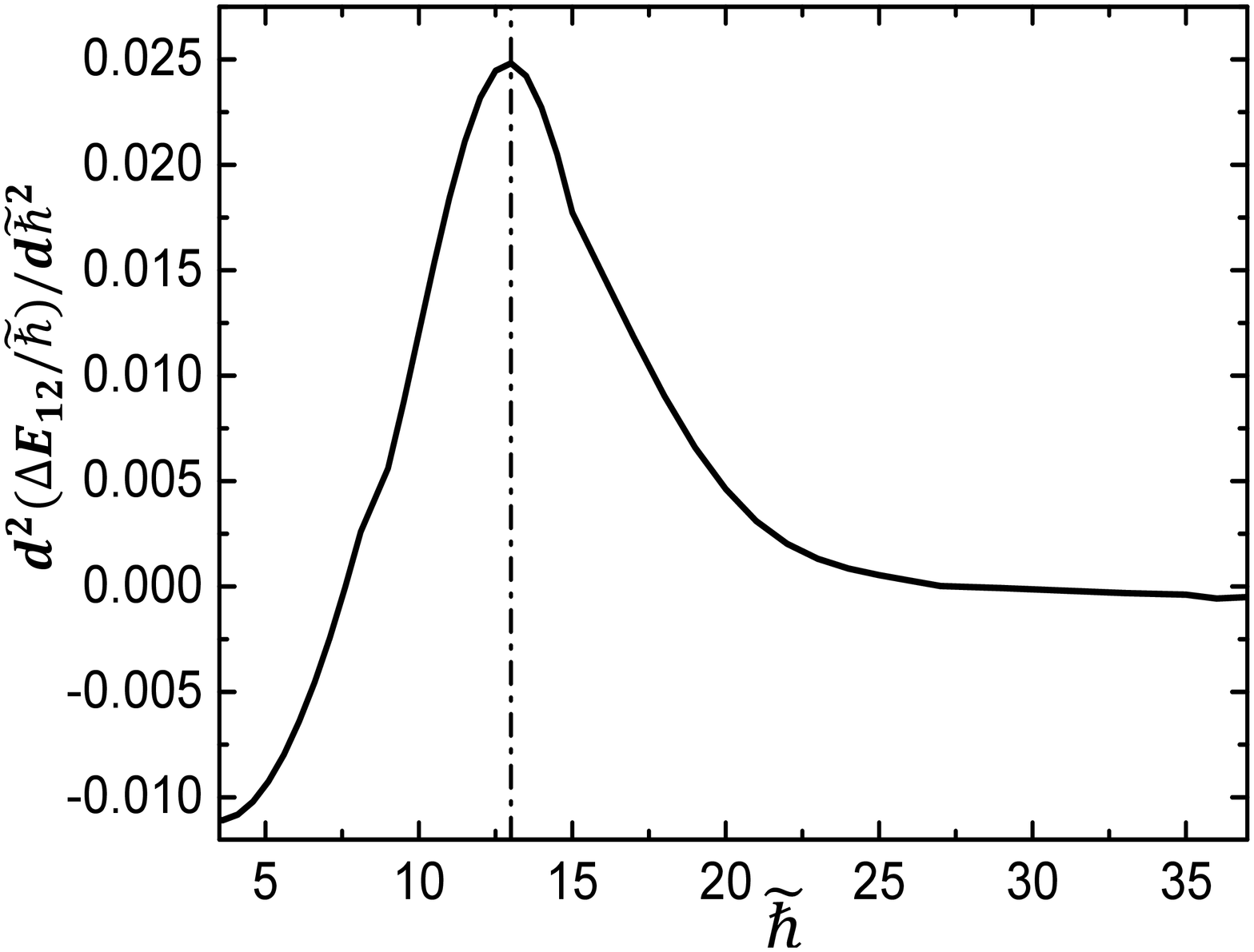}
\caption{ The   energy gap $\frac{\Delta E_{12}}{\tilde{\hbar}}$ and its   first and second  order derivatives of a FK chain. The  parameters are $N=100$ and $K=5$. }
\label{gap2}
\end{figure}

Fig. \ref{gap1} presents the result of the energy gap $\Delta E_{12}$ between the
ground state and the first excited state for different system sizes. Because the
energy gap for the harmonic oscillator chain, $ \Delta E_{12}=
\tilde{\hbar}\cdot \sin \frac{\pi}{2(N+1)}$, shows a linear dependence
upon $\tilde{\hbar}$, we have scaled the ordinate axis by $\tilde \hbar$ in Fig. \ref{gap1} so that $\Delta E_{12}/\tilde\hbar$ will become a constant in this ideal model.
It is interesting to note that when $\tilde\hbar$ is big enough, $\Delta E_{12}/\tilde\hbar$ in FK model also becomes a constant just like the harmonic oscillator chain. This result is
consistent with the conclusion from the above study upon the ground state energy. To get the transition point, we plot the second derivative of $\Delta E_{12}/\tilde\hbar$ with respect to $\tilde\hbar$ for the size $N=100$. Here again we find the characteristic value $\tilde{\hbar}_c\approx 13$, at which $\Delta E_{12}/\tilde\hbar$ shows a peak. So by summarizing all the data we have obtained until now, the phase transition point can be consitently set to be $\tilde{\hbar}_c\approx 13$.

\section{SUMMARY}

In summary, we have investigated in details the phase transition
from the pinned state to the slidingone in the one-dimensional
commensurate quantum FK model with a density-matrix renormalization
group algorithm. By looking into the entanglement, the particle
coordinate correlation, the ground state energy and the energy gap,
a transition point with $\tilde{\hbar}_c\approx 13$ is found when the
external potential is set to be $K=5$. In the sliding state, the
system behaves much like a chain of harmonic oscillators. It is
hoped that the research work in this paper will not only help us to
understand more the quantum FK model in describing many condensed
matter systems, but also find applications in quantum information
processing (QIP), such as the cold trapped ions in an optical
lattice, which has been put forward as a new realization of quantum
FK model recently \cite{Garcia}.
\par
This work is supported by the National Natural Science Foundation of
China under Grant Nos. 11274117 and 11134003 and Shanghai Excellent academic leaders Program of China (Grant No. 12XD1402400).

%
% BibTeX users please use
% \bibliographystyle{}
% \bibliography{}
%
% Non-BibTeX users please use

\end{document}